\input harvmac.tex
\vskip 0.5in
\Title{\vbox {\baselineskip 12pt
\hbox to \hsize{\hfill KEK-TH-767 }
\hbox to \hsize{\hfill HIP-2001-13/TH}
\hbox to \hsize{\hfill hep-th/0105055}}}
{\vbox {\centerline{Ghost Tachyon Condensation,
Brane-like States}
\vskip 0.1in
\centerline {and Extra Time Dimension}
}}
\centerline{{\bf Masud Chaichian} \footnote{$^1$}
{masud.chaichian@helsinki.fi}, {\bf Archil Kobakhidze}
\footnote{$^2$} {archil.kobakhidze@helsinki.fi}} \centerline{\it
High Energy Physics Division, Department of Physics, University
of Helsinki and} \centerline{\it Helsinki Institute of Physics,
FIN-00014 Helsinki, Finland} 

\smallskip

\centerline{\bf {Dimitri Polyakov} 
\footnote{$^3$} {polyakov@post.kek.jp, poliakov@ihes.fr }}
\centerline{\it High Energy Accelerator Research Organization
(KEK) Tsukuba, Ibaraki 305-0801 Japan}

\vskip .2in

{\baselineskip 12pt \centerline {\bf Abstract}{ NSR superstring  theory
contains a tower of physical  vertex operators (brane-like
states) which exist at non-zero pictures only, i.e. are
essentially mixed with superconformal ghosts. Some of these
states are massless, they are responsible for creating D-branes.
Other states are tachyonic (called ghost tachyons) creating a
problem for the vacuum stability of the NSR model. In this paper
we explore the role played by these tachyonic states in string
dynamics. We show that the ghost tachyons condense on D-branes,
created by massless brane-like states. Thus the vacuum stability
is achieved dynamically, as the effective ghost tachyon potential
exactly cancels the D-brane tension, in full analogy with Sen's
mechanism. As a result, from perturbative NSR model point of
view, massless and tachyonic brane-like  states appear to live in
a parallel world, as the brane is screened by the tachyonic veil.
We extend the analysis to the brane-antibrane pair in AdS space
and show that in this case due to the effect of 
the ghost tachyon condensation
one can construct extra time-dimensional phenomenological models
 without tachyons and antibranes.}
\Date{April 2001 {\bf PACS: $04.50.+h$;$11.25.Mj$.}} 
\vfill\eject
\lref\verlinde{E. Verlinde, H. Verlinde, JHEP {\bf 0005} (2000) 034.}
\lref\verlind{J. de Boer, E. Verlinde, H. Verlinde, JHEP {\bf 0008}
(2000) 003.} \lref\myself{D. Polyakov, Class. Quant. Grav. {\bf 18} (2001) 1979, hep-th/0005094.}
\lref\periwal{L.Lifschyts, V.Periwal, 
JHEP {\bf 0004} (2000) 026.}
\lref\ampr{A.M. Polyakov, Int. J. Mod. Phys. {\bf A14} (1999) 645.}
\lref\huffel{P. Damgaard, H. Huffel, Phys. Rep. {\bf 152} (1987) 227.}
\lref\ampf{S. Gubser, I. Klebanov, A.M. Polyakov, Phys. Lett. {\bf
B428} (1998) 105.} \lref\myselff{D. Polyakov, Phys. Rev. {\bf D57}
(1998) 2564.} \lref\parisi{G. Parisi, Y.S. Wu, Sci. Sinica {\bf 24}
(1981) 484} \lref\malda{J. Maldacena, Adv. Theor. Math. Phys. {\bf 2} (1998)
231.} \lref\wit{E. Witten, Adv. Theor. Math. Phys. {\bf 2} 
(1998) 253.} \lref\witt{E. Witten, Nucl. Phys. {\bf B276} (1986) 291.}
\lref\sen{A. Sen, JHEP {\bf 9808} (1998) 012.} \lref\senn{A. Sen, 
JHEP {\bf 9912} (1999) 027.} \lref\sz{A. Sen, B. Zwiebach, JHEP
{\bf 0003} (2000) 002.} \lref\szb{N. Berkovits, A. Sen, B. 
Zwiebach, Nucl. Phys. {\bf B587} (2000) 147.} \lref\nb{N. 
Berkovits, JHEP {\bf 0004} (2000) 022.} \lref\ck{M. Chaichian, A.B.
Kobakhidze, Phys. Lett. {\bf B488} (2000) 117.} \lref\rsh{V.A.
Rubakov, M.E. Shaposhnikov, Phys. Lett. {\bf B125} (1983) 136}
\lref\bran{M. Gogberashvili, hep-ph/9812296; L. Randall and R.
Sundrum, Phys. Rev. Lett. {\bf 83} (1999) 3370; I.I. Kogan, S.
Mouslopoulos, A. Papazoglou, G.G. Ross, J. Santiago, Nucl. Phys. 
{\bf B584} (2000) 313; I.I. Kogan, G.G. Ross, Phys. Lett. {\bf
B485} (2000) 255.}
\lref\kogan{I.I. Kogan, D. Polyakov, hep-th/0012128.}

\baselineskip 16pt

\centerline{\bf 1. Introduction}

\medskip

Recently it has been conjectured from string field theoretic
arguments that effective potential of the tachyonic mode of non
BPS Dp-branes in type II string theory has an important property
of universality, that is, the form of the tachyon potential,
bounded from below, remains the same for different Dp-brane
backgrounds and independent on specifics of boundary conformal
field theory ~{\sen, \senn}. An important argument in favour of
such a conjecture comes from open string field theory (OSFT);
Namely, the off-shell computation of the effective tachyon
potential in OSFT (truncated at the lowest massive level) shows
that the OSFT tachyon potential is bounded from below; its value
at its minimum  is negative and precisely cancels the non BPS
Dp-brane tension (or the tension of brane-antibrane pair). As a
result of such a screening of D-branes, caused by the
condensation of tachyon at minimum point of the potential, one
effectively obtains the type II flat vacuum without branes, with
full ten-dimensional Lorentz symmetry restored ~{\sen, \senn}. In
standard evaluations of the tachyon potential in open superstring
field theory one usually imposes strict constraints on the form
of possible off-shell CFT vertex operators that contribute to the
effective OSFT Wess-Zumino type action. In particular, one  fixes
the picture-changing gauge symmetry by restricting the vertices
to have picture 0 (with the picture/ghost number assignment
adopted in ~{\szb}); such a gauge fixing is needed to avoid
summing over physically identical OSFT modes (which differ by a
picture changing
 transformation). So one starts with the tachyonic state
at the lowest level of OSFT; consequently, the higher level
states are constructed by acting on the tachyon vertex operator
with various combinations os stress tensor, supercurrent and
ghost fields ~{\szb}. Such a consistent truncation of the OSFT is
sufficient to determine the effective potential  of tachyonic
mode in the Dirichlet strings in flat background. At the same
time, this derivation does not take into account the OSFT modes
which exist at non-zero pictures only, but which do play a
significant role in the open string dynamics. In  NSR critical
string theory  these vertex  operators correspond to a  specific
class of NS physical states (BRST invariant and BRST nontrivial
vertex operators) which are not associated with any perturbative
particle emission but appear to play an important role in the
non-perturbative physics. For example in the previous works it
has been shown that inserting   this new class of operators to NSR
 string theory is
 equivalent to introducing branes;
In particular, some of these operators dynamically deform flat
ten-dimensional space time to $AdS_5\times{S^5}$ background
~{\myself}.

Generally, these states are described by vertex operators
that exist at nonzero ghost pictures only, not admitting ghost
number zero representation. This crucially distinguishes them from
usual perturbative open or closed string states, such as
photon, graviton or dilaton, which in principle are allowed to
exist at any arbitrary ghost picture.
This new class of states appears in both closed and open superstring
theories and includes both massless and tachyonic states.
In open string theory the massless ones are represented by two-form
and five-form vertex operators;
they are given by:

\eqn\grav{\eqalign{V_{m_1...m_5}^{(-3)}(z,k)=
e^{-3\phi}\psi_{m_1}...\psi_{m_5}e^{ikX} \cr
V_{m_1...m_5}^{(+1)}(k)=e^{\phi}\psi_{m_1}...\psi_{m_5}e^{ikX}(z)
+ghosts\cr
V_{m_1m_2}=e^{-2\phi}\psi_{m_1}\psi_{m_2}e^{ikX}}}
These dimension 1 primary fields must also be integrated over the
worldsheet boundary (or multiplied by the fermionic ghost field $c$).
Apart from the massless (on shell) 5-form vertex operator
one can also construct the tachyonic 4-form state in the $-3$-picture given by
\eqn\lowen{V_{m_1...m_4}=e^{-3\phi}\psi_{m_1}...\psi_{m_4}{e^{ikX}}(z)}
as well as the two-form:
\eqn\lowen{V_{ab}=e^{-3\phi}\partial{X_a}\partial{X_b}e^{ikX}}
As is easy to see, the on-shell condition for this state is given by
$k^2=-1$
Continuing this way, one may also construct
 tachyonic states of bigger negative $k^2$
in the $-3$-picture, i.e. the ,$1-$form, the $2-$form and the  $3-$form,
corresponding to
the on-shell masses $k^2=-4,-3$ and $-2$.
Many  important
questions naturally arise with regard to these new open string massless
and tachyonic states. As was stressed above, the 5-form vertex does not
correspond to any perturbative open string mode, but is rather related to
dynamics of branes. However, doesn't this additional physical state in
the  NSR superstring spectrum lead to violation of
 unitarity of open string
amplitudes? On the other hand, as for the $-3$-picture tachyonic states,
their appearance clearly may create a problem for the vacuum stability.
At the same time, in principle there is no reason to exclude
these extra open string states from consideration, as  they all
appear to be physical - i.e. BRST invariant and BRST non-trivial,
and the only reason why they were overlooked before is because
of their mixing with ghosts - there is no picture zero representation for
these operators.
So clearly it is of interest to investigate the non-perturbative dynamics
related to the brane-like massless and tachyonic states and
in particular to understand how to reconcile their appearance
with unitarity and vacuum stability in string theory.
In the present paper we shall attempt to address these questions
showing that in fact the massless and tachyonic
brane-like states ``compensate'' each other due to
mechanism similar to the one proposed in
 the Sen's conjecture. Namely, as has been mentioned
above, introducing the 5-form vertex operator is equivalent
 introducing a D-brane with a certain mass (tension) to the theory;
on the other hand, the tachyonic brane-like states (2) condense
on the brane created by the 5-form, just like in the context of
Sen's conjecture usual tachyon condenses on a D-brane. Then we
show, through  open string field theory calculations, that the
negative energy density of the condensate of the
 brane-like tachyon (2) (which we also
refer to as the ghost tachyon) at the minimum of the effective potential
precisely cancels the positive tension of the brane created by the
 5-form vertex operator (1). This phenomenon is exactly analogous to
what happens in the framework of Sen's mechanism and we call it
``parallel tachyon condensation''.
The phenomenon of parallel tachyon condensation actually resolves
both stability and unitarity problems at the same time:
as the tachyon gets localized on the brane (related to the 5-form),
the energy density at the potential minimum screens the positive
brane tension and therefore the brane becomes ``invisible'' from
the usual perturbative string theory point of  view - and such a
screening effectively resolves the unitarity issue.
In fact, it appears as if the 5-form brane-like states live
in a parallel world, invisible from perturbative string theory point of view.

This paper is organized as follows. In the section 2, we review
the $on-shell$ BRST properties of the brane-like states to show 
their invariance and non-triviality under BRST transformations -
to stress their importance to open string and D-brane dynamics
(for detailed analysis see ~{\kogan}).
In the sections 3,4,5, we develop the string field theory network
to compute the effective ghost tachyon potential and also
calculate the brane tension associated with massless brane-like
vertex operators. In the section 6 we calculate the effective
ghost tachyon potential and analyze its mimimum. We show, by
comparing the brane-tension and the minimum value of the
potential, that the analogue of the the Sen's conjecture does
emerge in this case and the potential minimum cancels the brane
mass. In the section 7 we will  discuss the relation between the
ghost tachyon vertex operators and tachyonic solutions for the
equations of motion in theories with time-like extra dimension.
We attempt to interpret the ghost tachyonic vertex operators (2),
(3) as emission vertices for tachyonic KK modes which appear as
the e.o.m. solutions in the above theories and analyze the
effective potential of these tachyonic modes  in the context of
brane world scenario with extra time-like dimension. Usually
brane world scenarious are known to involve branes
 with negative tensions - which are actually  fictitious   and unnatural
objects because of their instability. We argue, however, that
 tachyonic solutions in theories with extra time-like dimension
can cure the antibrane instability as
the effective potential for ``tachyonic gravitons'' is bounded from below and
 the tachyonic gravitons
condense on the antibrane (with $positive$ energy density
at the potential minimum) and cancels the negative tension of
the antibrane insuring its stability. As a result, due to such
an ``anti-Sen'' scenario, one can consider phenomenological
theories with extra time-like dimensions involving stable brane
configurations: negative tension of a hidden brane is screened by the
condensate of the tachyonic KK modes - the e.o.m. solutions
(in the meantime the visible non-BPS brane with positive tension is to be
 stabilized by the usual tachyon condensation scenario). Finally,
 in the concluding section 8 we discuss some other possible
 implications of our results.

\medskip

\centerline{\bf 2. Brane-like states - review of BRST properties}

\medskip

The five-form picture $-3$ operator $V^{(-3)}_{m_1...m_5}$ is
BRST-invariant at any momentum, as is easy to see by simple and
straightforward computations involving the BRST charge; however
it is BRST non-trivial only if its momentum k is polarized along
the 5 out of 10 directions orthogonal to $m_1$,...$m_5$ (for any
given polarization choice of $m_i$). Indeed, it is easy to see
that the only BRST triviality threat for the 5-form operator may
appear from the expression
$\lbrace{Q_{brst},S_{m_1...m_5}}\rbrace$ where
$S_{m_1...m_5}=e^{\chi-4\phi}\partial\chi(\psi\partial{X})
\psi_{m_1}...\psi_{m_5}e^{ikX}$ - indeed, its commutator with the
$\oint{\gamma(\psi\partial{X})}$ in the BRST  charge does give the
$V_5$ operator, while the $\oint{\gamma^2b}$ of $Q_{brst}$
commutes with S. However, it is clear that if the momentum $k$ is
orthogonal to $m_1,...m_5$ directions the $S_{m_1...m_5}$ is not
a primary field: the $(\psi\partial{X})$ part of it always has
internal O.P.E. singularities with either
$\psi_{m_1}...\psi_{m_5}$ or $e^{ikX}$. As a result, whenever the
momentum is orthogonal to $m_1,...m_5$ directions, the O.P.E. of
the stress-energy tensor with $S_{m_1...m_5}$ always has a cubic
singularity. Therefore $S_{m_1...m_5}$ does not commute with the
$\oint{cT}$ term of $Q_{brst}$ and
$\lbrace{Q_{brst},S_{m_1...m_5}}\rbrace$ does not reproduce the
5-form vertex operator $V^{(-3)}_{m_1...m_5}$. However, in case if
the momentum $k$ of $V^{(-3)}_{m_1...m_5}$ is longitudinal, i.e.
is polarized along $m_1...m_5$ directions it is easy to see that
the vertex operator becomes BRST trivial: indeed, it can be
written as a BRST commutator with the primary field:
$\lbrace{Q_{brst},C_{m_1...m_5}}\rbrace$ where
$C_{m_1...m_5}=e^{\chi-4\phi}\partial\chi(\psi\partial{X})^{\perp}
\psi_{m_1}...\psi_{m_5}e^{ikX}$ with the supercurrent part
$(\psi\partial{X})^{\perp}$ now polarized orthogonally to
$m_1,...m_5$, i.e.
 both to $e^{ikX}$ and other worldsheet fermions
of the 5-form. So we see that BRST non-triviality condition
imposes significant constraints on the propagation of the
5-form: namely, it is allowed to propagate in the
5-dimensional subspace transverse to its own polarization.
This also is a remarkable distinction of this vertex operator
from usual vertices we encounter in perturbative string theory;
it is well known that those are able to propagate in entire
ten-dimensional space-time.
The two-form is also BRST-invariant at any k;
although it is BRST-trivial at zero momentum as it can be represented as
a commutator
$\lbrace{Q_{brst},e^{\chi-3\phi}\psi_{\lbrack{m_1}}\partial{X}_{m_2{\rbrack}}
+e^{\chi-3\phi}(\psi\partial{X})\psi_{m_1}\psi_{m_2}
\rbrace}$,  it becomes BRST non-trivial at non-zero momenta
and again, in complete analogy with the 5-form case,
its momentum  must be orthogonal to the $m_1,m_2$ two-dimensional
subspace, i.e. the two-form propagates in eight transverse dimensions.

Constructing the BRST-invariant version of the five-form
at picture $+1$
is a bit more tricky since the straightforward generalization
given by  $e^\phi\psi_{m_1}...\psi_{m_5}$
does not commute with two terms in the BRST current
given by $b\gamma^2$ and $\gamma{\psi_m\partial{X^m}}$.
To compensate for this non-invariance one has to add  two
counterterms, one proportional to the fermionic
ghost number 1 field ${c}$ and another to the ghost
number $-1$ field b.

To construct these ghost counterterms one has
to take the fourth power of picture-changing operator
$\Gamma^4\sim{:e^{4\phi}G\partial{G}\partial^2G\partial^3G:}$
with G being the full matter $+$ ghost worldsheet supercurrent
and calculate its full O.P.E. (i.e. including all the non-singular terms)
 with the picture $-3$ five-form operator.
If the $-3$-picture vertex operator is at the point 0 then
\eqn\grav{\eqalign{
V_5^{(+1)}(0)=e^{\phi}\psi_{m_1}...\psi_{m_5}-{1\over2}
lim_{z\rightarrow{0}}{\lbrace}{z^2}be^{2\phi-\chi}\psi_{m_1...m_5}\cr-{1\over2}
{z^2}ce^\chi\psi_{m_1}...\psi_{m_4}(\psi_{m_5}(\psi_n\partial{X^n})
+\partial{X_{m_5}}(\partial\phi-\partial\chi))+
O(z^3)\rbrace}}
This  operator is  BRST invariant by construction since both
$\Gamma^4$  and picture $-3$ 5-form operator are BRST invariant.

The BRST commutator with counterterms must be computed at a point $z$
and then the limit $z\rightarrow{0}$ is to be taken.
Fortunately, due to the condition of fermionic ghost number conservation
this unpleasant non-local ghost part is unimportant in
computations of correlation functions and can be dropped at least
in cases when not more than one picture $+1$-operator is involved.
Also, since in open string field theory all the vertices  are off-shell,
the BRST invariance is not important and therefore we shall consider only the
local part of the $+1$-picture 5-form in all our OSFT computations.

\medskip

\centerline{\bf 3. Effective potential of the ghost tachyon}

\medskip

In this section we will consider the contribution of the massless
tachyonic brane-like vertices (2),(3) to the OSFT effective
action. Let us start with reviewing the basic framework of open
superstring field theory (for full details see ~{\sz, \szb, \nb}).
The off-shell superstring field theory is described by the
following  Wess-Zumino type action
\eqn\grav{\eqalign{S={1\over{2g^2}}<<(e^{-\Phi}{Q_{brst}}e^{\Phi})(e^{-\Phi}
{\eta_0}e^{\Phi})-\int_0^1{dt}(e^{-t\Phi}\partial_t{e^{t\Phi}}){\lbrace}(e^{-t\Phi}Q_{brst}e^{t\Phi})
(e^{-t\Phi}\eta_0{e^{t\Phi}})\rbrace>>}}
where
$b,c,\beta,\gamma$ are fermionic and superconformal ghosts with bosonization
formulae
\eqn\grav{\eqalign{c=e^\sigma,b=e^{-\sigma}\cr
\gamma=e^{\phi-\chi},\beta=e^{\chi-\phi}\partial\chi\cr
<\sigma(z)\sigma(w)>=<\chi(z)\chi(w)>=-<\phi(z)\phi(w)>=log(z-w)\cr
\eta(z)\equiv{e^{-\chi}}(z),\xi(z)\equiv{e^\chi}(z)}}
and BRST operator is given by
\eqn\lowen{Q_{BRST}=\oint{{dz}\over{2i\pi}}(c(T_m+T_{gh})+{1\over2}e^{\phi-\chi}
\psi_m\partial{X^m}+{1\over4}e^{2\phi-2\chi}b-b:c\partial{c})}
The time ordered correlators $<<...>>$ are defined by:
\eqn\grav{\eqalign{<<V_1...V_N>>=<f_1^{(N)}\circ{V_1(0)}...f_N^{(N)}
\circ{V_N(0)}>\cr
f_k^{(N)}(z)=e^{{{2i\pi(k-1)}\over{N}}}{({{1+iz}\over{1-iz}})}^{{2\over{N}}}}}
where the function f(z) defines the conformal transform of V by f,
for instance, if $V$ is a primary field of dimension h, one has
$f\circ{V(0)}=({{df}\over{dz}})^h{V(f(0))}$ This action is
invariant under the local gauge transformation ~{\nb} given by
\eqn\lowen{\delta{e^\Phi}={(Q_{{brst}}\epsilon)}e^{\Phi}+e^{\Phi}(\eta_0\lambda)}
where $\epsilon,\lambda$ are Grassmann odd. Using this gauge
symmetry it is convenient to choose the gauge
\eqn\lowen{b_0\Phi=0, \xi_0\Phi=0} String fields satisfying these
conditions are related to local NS operators by ~{\witt}
\eqn\lowen{\Phi=:\xi{V}:}
For example, if $V=ce^{-\phi}$ is tachyon vertex operator in the $-1$-picture,
the corresponding string field will be given by the following picture
zero operator: $\Phi=ce^{\chi-\phi}$
Also, it should be noted that, since all the GSO(-) states are Grassmann,
odd,
they should be accompanied by the appropriate $2\times{2}$ internal Chan-Paton
factor, i.e. the Pauli matrix $\sigma_1$. At the same time, the GSO(+)
 fields must be multiplied by the $2\times{2}$ identity matrix
so that the complete string field may be written as
\eqn\lowen{{\hat\Phi}=\Phi_{-}\otimes\sigma_1+\Phi_{+}\otimes{I}}
Also,  ${\hat{Q}}_{brst}=Q_{brst}\otimes{\sigma_3}$
and ${\hat\eta_0}=\eta_0\otimes{\sigma_3}$

It is convenient to expand the exponents of the OSFT action (5)
in power series, so that the action is expressed as
\eqn\lowen{S={1\over{2g^2}}{\sum_{M,N=0}^{\infty}}
{{{(-1)^N}(M+N)!}\over{M!N!(M+N+2)!}}<<({\hat{Q}}_{brst}{\hat\Phi}){{(\hat\Phi)}^M}({\hat\eta_0}{\hat\Phi}){({\hat\Phi})^N >>}}
To compute the effective potential of the tachyonic modes (2),(3)
one has to specify the consistent truncation ~{\sen, \senn} of
the string field $\Phi$ which takes into account the brane-like
vertices. Let $H_{m,n}$ be a subspace of operators of ghost
number m and picture n. First of all, as in ~{\szb} the
truncation we are looking for will  include all the operators in
the
 $H_{0,0}$-subspace of total picture zero and total ghost number zero;
As has been shown in ~{\szb}, the $Z_2$ twist-even string field,
satisfying the gauge condition (11)  and restricted to $H_{0,0}$
up to level $3/2$ is given by
\eqn\grav{\eqalign{{\hat{\Phi_{0,0}}}=t{\hat{T}}+a{\hat{A}}+e{\hat{E}}
+f{\hat{F}}\cr
{\hat{T}}=\xi{c}e^{-\phi}\otimes{\sigma_1}\cr
{\hat{A}}=c\partial^2{c}\xi\partial\xi{e^{-2\phi}}\otimes{I}\cr
{\hat{E}}=\xi\eta\otimes{I}\cr
{\hat{F}}={-1\over2}\xi(\psi_m\partial{X^m})\otimes{I}}}
However, in our case, to account for the brane-like states,
 the consistent truncation
  also  must include operators of the subspaces $H_{0,2}$ and $H_{0,-2}$
which do not admit a picture zero representation; let us denote
the appropriate subspaces as ${\tilde{H}}_{0,2}$ and
${\tilde{H}}_{0,-2}$ ; of course,
${\tilde{H}}_{0,2}\subset{H_{0,2}}$ and
${\tilde{H}}_{0,-2}\subset{H_{0,-2}}$. At the lowest levels, the
subspace $\tilde{H}_{0,-2}$ consists of the following vertex
operators: $ce^{\chi-3\phi},
 ce^{\chi-3\phi}\psi_{m_1}...\psi_{m_p}(p=1,..5),
 ce^{\chi-3\phi}\partial{X_m},
 ce^{\chi-3\phi}\partial{X_m}\partial{X_n}$ and the same for
$\tilde{H}_{0,2}$ with $e^{\chi-3\phi}\rightarrow{e^{\chi+\phi}}$
At the same time, the operators
\eqn\grav{\eqalign{{\hat{P^{(-3)}}}=ce^{\chi-3\phi}(\psi_m\partial{X_m})
\otimes{I}\cr
{\hat{Q_{np}^{(-3)}}}=ce^{\chi-3\phi}(\psi_m\partial{X^m})\psi_n\psi_p
\otimes{I}\cr
{\hat{P^{(+1)}}}=ce^{\chi+\phi}(\psi_m\partial{X_m})\otimes{I}\cr
{\hat{Q_{np}^{(+1)}}}=ce^{\chi+\phi}(\psi_m\partial{X^m})\psi_n\psi_p
\otimes{I}}}
do $not$ belong to ${\tilde{H}}_{0,-2}$ (and also ${\tilde{H}}_{0,2}$)
since ${\hat{P}}^{(-3)}$ and ${\hat{Q}}_{np}^{(-3)}$ can be transformed
to higher picture representation by the picture changing transformation
and
${\hat{P}}^{(+1)}$ and ${\hat{Q}}_{np}^{(+1)}$ are
nothing but unity operator and Lorentz generator in the $+1$-picture
(multiplied by $c\xi$)
So we will be looking for the truncation
of the OSFT that includes the vertices from the above subspaces
$H_{0,0}$, ${\tilde{H}}_{0,2}$ and ${\tilde{H}}_{0,-2}$
and which is consistent up to level ${3\over2}$ of the open superstring
 field theory (i.e. the same maximum level considered in
~{\szb}). Moreover, not all of the operators of
${\tilde{H}}_{0,2}$ and ${\tilde{H}}_{0,-2}$ must be included in
the mode expansion of the open string field but only those which,
in a group with the vertices of $H_{0,0}$ define a consistent
truncation of the open string field theory (so that the effective
OSFT equations of motion could be obtained by merely varying the
OSFT action with respect to the vertices belonging to the
truncation set). The consistency of the OSFT truncation is
equivalent to the
 following condition. Suppose $\cal{H}$ is a truncation.
Then,  for the truncation to be consistent, the OSFT action must
always be quadratic or higher order in the components of $\Phi$
orthogonal to $\cal{H}$. Then, if one takes $\Phi$ to be entirely
inside $\cal{H}$, the OSFT equations of motion, obtained by
varying $S$ with respect to all the states of
${{\lbrace{\Phi}\rbrace}\over{\cal{H}}}$ are satisfied
automatically and the truncation is consistent ~{\senn}. It is
now not difficult to check that the truncation consistency
condition (up to level ${3\over2}$ excludes the following
tachyonic $\tilde{H}_{0,-2}$ operators:
${\hat{R}}=ce^{\chi-3\phi}\otimes{\sigma_1}$,
${\hat{R}_{np}}=ce^{\chi-3\phi}\psi_{n}\psi_{p}\otimes{\sigma_1}$
and ${\hat{R}_m}=ce^{\chi-3\phi}\partial{X_m}\otimes{\sigma_1}$
from
 from $\cal{H}$  because the following OSFT correlators:
\eqn\grav{\eqalign{<<({\hat{Q}}_{brst},{\hat{T}}){\hat{R}}({\hat\eta}_0
{\hat{P}^{(-3)}}){\hat{F}}>>,\cr
<<({\hat{Q}}_{brst},{\hat{T}}){\hat{R}_{np}}({\hat\eta}_0
{\hat{P}_{np}^{(-3)}}){\hat{F}}>>,\cr
<<({\hat{Q}}_{brst},{\hat{T}}){\hat{R}_m}({\hat\eta}_0
{\hat{P}_m^{(-3)}}){\hat{F}}>>}} do not vanish and therefore the
OSFT action has the terms linear in ${\hat{P}}_{}^{(-3)}$,
${\hat{P}}_{m}^{(-3)}$ and ${\hat{P}}_{np}^{(-3)}$ unless we
exclude ${\hat{R}}_{}$, ${\hat{R}}_{m}$ and ${\hat{R}}_{np}$ from
the truncation scheme. The totally similar argument excludes also
the analogues of  ${\hat{R}}_{}$, ${\hat{R}}_{m}$ and
${\hat{R}}_{np}$ carrying the ghost $\phi$ number $+1$. Next, as
in ~{\szb} the OSFT action truncated on $\cal{H}$ has a twist
$Z_2$-symmetry under which string fields associated with vertices
of dimension $h$ carry charge $(-1)^{h+1}$ for even $2h$ and
$(-1)^{h+{1\over2}}$ for odd $2h$. Using this twist symmetry we
can further truncate the string field components from the
 ${\tilde{H}_{0,2}}$ and ${\tilde{H}_{0,-2}}$ sectors to exclude the
brane-like tachyonic states:

$ce^{\chi-3\phi}\psi_m$,

$ce^{\chi-3\phi}\psi_{m_1}\psi_{m_2}\psi_{m_3}$,

$ce^{\chi-3\phi}\psi_{m_1}...\psi_{m_5}$,

$ce^{\chi+\phi}\psi_m$

$ce^{\chi+\phi}\psi_{m_1}\psi_{m_2}\psi_{m_3}$;

$ce^{\chi+\phi}\psi_{m_1}...\psi_{m_5}$,

so that only the tachyonic 4-form brane-like string field is left
in the expansion. In the rest of the paper, for the sake of
certainty, we will be analyzing the case of tachyon condensation
on the D3-brane. Finally we are ready to write the relevant
expansion for consistently truncated OSFT needed to evaluate the
ghost tachyon potential:
\eqn\grav{\eqalign{{\hat{\Phi}}=t{\hat{T}}+a{\hat{A}}+e{\hat{E}}
+f{\hat{F}}+\tau{\hat{\Theta}}+\lambda_{ab}{\hat{\Lambda}_{ab}}\cr
{\hat{T}}=\xi{c}e^{-\phi}\otimes{\sigma_1}\cr
{\hat{A}}=c\partial^2{c}\xi\partial\xi{e^{-2\phi}}\otimes{I}\cr
{\hat{E}}=\xi\eta\otimes{I}\cr
{\hat{F}}={-1\over2}\xi(\psi_m\partial{X^m})\otimes{I}\cr
{{\hat{\Lambda}}_{ab}}=c\xi(e^{\phi}+e^{-3\phi})\partial{X_a}\partial{X_b}\cr
{\hat{\Theta}}=c\xi(e^{\phi}+e^{-3\phi})\psi_{a_1}...\psi_{a_4}
\epsilon^{a_1...a_4}\cr a,b,a_i=0,...3}} In the last formula we
have chosen the polarizations of the ghost tachyon so that the
4-form spans the 4-dimensional $D3$-brane worldvolume; the
tachyonic graviton is localized in the D3-brane worldvolume as
well. Therefore in our construction the ghost tachyonic modes
consist of the tachyonic scalar (corresponding to the ghost
tachyonic 4-form) and the 4-dimensional tachyonic 2-form
(graviton) Also, to determine the contribution of brane-like
states to the modified $D3$-brane tension (which is to be compared
with the energy density at the minimum of the ghost tachyon
potential), we shall need to properly polarize the massless
$5$-form:
\eqn\lowen{\Lambda_j=c\xi(e^{\phi}+e^{-3\phi})\psi_0\psi_1\psi_2\psi_3\psi_j
(j=4,...,9)}
i.e. 4 out of five fermions are directed along the $D3$-brane
worldvolume while the fifth is the ``orthogonal'' one, accounting
for the transverse oscillation of the $D3-brane$, associated with
the $5$-form brane-like state.

\medskip

\centerline{\bf 4. Brane-like states and D3-brane tension}

\medskip

In this section we shall calculate the tension of the brane
associated with the massless $5$-form vertex operator and show
that it is equal to the tension value for   the usual $D3$-brane,
i.e. $1\over{2\pi^2{g^2}}$ To compute the tension, corresponding
to the massless 5-form, we use the strategy totally similar to
~{\sen, \szb}. Namely,
\eqn\grav{\eqalign{{\hat{V}}_5=\lambda_j(k)
ce^{\chi}(e^{\phi}+e^{-3\phi})\psi_0...\psi_3\psi_j
e^{ik_i{X^i}}(\tau)\cr
i,j=4,...9}}
one has to consider the  $k$-dependent  quadratic term involving
the $\lambda_j$ mode in the OSFT action (5), given by
${1\over{2g^2}}<<(Q_{brst}{\hat{V}}_5)(\eta_0{\hat{V}}_5)>>$. The
only term in the BRST charge contributing to the $k$-dependent
part of the quadrtic term is given by
$\oint{{d\tau}\over{2i\pi}}cT_{matter} (\tau)$. Evaluating its
commutator with ${\hat{V}}_5$ we find the kinetic part of the
quadratic term in the effective Lagrangian to be given by
\eqn\lowen{L^{(\lambda)}_{kin}=
{1\over{g^2}}\sum_{k}(k)^2\lambda_j(k)\lambda^j(-k)}
or, after the Fourier transform,
\eqn\lowen{L^{(\lambda)}_{kin}={1\over{g^2}}\partial_t{\lambda_j}
\partial_t{\lambda^j}}
Now, to find a tension of the brane one has to find the relation
between $\lambda_j$ and transverse collective coordinates
$\Lambda_j$ of the D-brane associated with $V_5$. As in ~{\szb},
$\lambda_j=\alpha\Lambda_j$ up to normalization factor. Once we
find the value of this normalization constant, we can elucidate
the brane tension, as the overall normalization of the kinetic
term, written in terms of the $\Lambda_j$ collective coordinates
is equal to the half of the square of the brane mass. In order to
find the normalization constant $\alpha$ one has to consider the
brane-antibrane pair (created by the $V_5$ vertices with
appropriate external Chan-Paton's factors) at a distance $b^t$
along  $X^t$ transverse direction, and an open string stretched
between them which has the tension given by ${1\over{2\pi}}
|\vec{b^t}|$. If one moves one of the branes by an amount $Y^t$
along $X^t$, the tension (or the mass$)^2$ change, up to the
first order in $Y$ is given by
\eqn\lowen{{1\over{(2\pi)^2}}(b_t{Y^t})}
On the other hand, just as in the standard D-brane case,
shifting one of the branes along the $X^t$ direction corresponds
to turning on the string field background given by
$ce^{\chi-\phi}\psi^t\otimes{I}\otimes
{\left[\matrix{1&0\cr{0}&{0}}\right]}$.
Therefore in order to find the normalization constant
$\alpha$ one has to compare this tension change
with the one that follows from relevant three-point
OSFT correlation functions involving one photon vertex
operator (corresponding to the open string between the brane and the antibrane)
and two massless five-form vertices (with necessary
external CP factors) that account for the brane and the antibrane.
The relevant string field expansion is given by:
\eqn\grav{\eqalign{\hat{\Phi}=\xi_t{c}{\hat{P}}^t
+\lambda_t(k){\hat{U}}_5^t(k)+\lambda^{*}_t(k){\hat{W}}_5^t(k)\cr
{\hat{P}}^t=ce^{\chi-\phi}\psi^t\otimes{I}\otimes
{\left[\matrix{1&0\cr{0}&0}\right]}
{\hat{U}}^t(k)=ce^\chi(e^\phi+e^{-3\phi})\psi_0...\psi_3\psi^t{e^{i
({\vec{k}}+{{\vec{b}}\over{2\pi}}){\vec{X}}}}\otimes{I}\otimes
{\left[\matrix{0&1\cr{0}&0}\right]}\cr
{\hat{W}}^t(k)=ce^\chi(e^\phi+e^{-3\phi})\psi_0...\psi_3\psi^t{e^{i
({\vec{k}}-{{\vec{b}}\over{2\pi}}){\vec{X}}}}\otimes{I}\otimes
{\left[\matrix{0&0\cr{1}&0}\right]}}}
Our aim now is to establish connection between the $\lambda^t$
polarization vector of the 5-form and the  transverse collective
coordinates $\Lambda^t$ of the D-brane created by the 5-form
vertex. Finding the normalization constant relating $\lambda$ and
$\Lambda$ will allow us to determine the tension of the brane.
This normalization must be determined from the
$({\vec{b}}{\vec{\xi}})({\vec{\lambda}}{\vec{\lambda^{*}}})$
coupling  originating from relevant three-point correlators in
string field theory involving $\hat{P}$, $\hat{U}$ and $\hat{W}$.
The three point OSFT vertex is given by
\eqn\grav{\eqalign{{1\over{12g^2}}(<<({\hat{Q}}_{brst}{\hat{\Phi}})
{\hat{\Phi}}({\hat{\eta}}_0{\hat{\Phi}})>>-
<<({\hat{Q}}_{brst}{\hat{\Phi}})({\hat{\eta}}_0{\hat{\Phi}})
{\hat{\Phi}}>>)}}
It is easy  to see that this vertex involves two
different classes of three-point correlators, those involving the
commutator of BRST charge with the photon ${\hat{P}}$ and those
with the commutators of ${\hat{Q}}_{brst}$ with the 5-forms
${\hat{U}}$ and ${\hat{W}}$. Using the expression (7) for
$Q_{brst}$ we find that these BRST commutators are given by:
\eqn\lowen{\lbrace{\hat{Q}}_{brst},{\hat{P}}^t\rbrace=
c\partial{X^t}\otimes{\sigma_1}\otimes{\left[\matrix{1&0\cr{0}&0}\right]}}
As for the BRST commutators with the five-forms ${\hat{U}}$ and
${\hat{W}}$ , we are only interested in the commutators of
$\int{{dz}\over{2i\pi}}(\gamma{G_{matter}})$ of
${{\hat{Q}}_{brst}}$ with ${\hat{U}}$ and ${\hat{W}}$ as these
are the only commutators contributing to three-point correlation
functions (since these are the only ones proportional to the
ghost field $c$). So we have
\eqn\grav{\eqalign{-{1\over2}\lbrace\int{{dz}\over{2i\pi}}e^{\phi-\chi}
\psi_m\partial{X^m}\otimes{\sigma_1}\otimes{I},
\lambda_t(k){\hat{U}}^t(k)\rbrace\cr=\lambda_t(k)\lbrack
\lbrace{c}e^{2\phi}\psi_0...\psi_3\psi^t\lbrack
{G}_{matter}P^{(1)}_{\phi-\chi}+\partial{G}_{matter}\rbrack{e^{ikX}}
\cr-{1\over2}\lambda_t(k)ce^{2\phi}\lbrack\psi_{\lbrack{0}}...\psi_2\partial
{X_{3\rbrack}}P^{(2)}_{\phi-\chi}\cr+\psi_{\lbrack{0}}...\psi_2\partial^2
{X_{3\rbrack}}P^{(1)}_{\phi-\chi}+{1\over2}\psi_{\lbrack{0}}...\psi_2\partial^3
{X_{3\rbrack}}\rbrack\psi^t{e^{ikX}}\cr
-{1\over2}\lambda_t(k)ce^{2\phi}\lbrack\psi_0...\psi_3{e^{ikX}}
\lbrack(\partial{X^t}+i(k\psi)\psi^t)P^{(2)}_{\phi-\chi}\cr+
(\partial^2{X^t}+i(k\partial\psi)\psi^t)P^{(1)}_{\phi-\chi}+
{1\over2}(\partial^3{X^t}+i(k\partial^2\psi)\psi^t)\rbrack+...\rbrack
\otimes{I}\otimes{\left[\matrix{0&1\cr{0}&0}\right]}}}
and the same formula for $\lbrace{\hat{Q}}_{brst},{\hat{W}}\rbrace$,
with only the external CP factor changed.
In the formula (26) the square brackets encompassing
the longitudinal indices $0,1,2,3$ denote the antisymmetrization
over these indices; the $P_{\phi-\chi}^{(1),(2)}$ are
the polynomials in derivatives of free $\phi$ and $\chi$ fields
(of conformal weights 1 and 2 respectively)
that emerge as a result of the differentiation of
the $\gamma(z)=e^{\phi-\chi}(z)$ ghost field with respect to $z$.
Also, we have dropped all the terms that are irrelevant
(i.e. not contributing) to the
$({\vec{\xi}}{\vec{b}})({\vec{\lambda}}{\vec{\lambda^{*}}})$
coupling that we seek
to elucidate from three-point string field correlation functions.
For instance, we are not interested in
couplings involving
the scalar product $\lambda$ and $\xi$ and all the terms in
three-point correlators giving rise to these couplings will be
dropped for the sake of shortness.
Finally, note that a half of all the three-point correlation functions vanish due
to the matrix identity
\eqn\lowen{Tr({\left[\matrix{0&0\cr{1}&{0}}\right]}
{\left[\matrix{0&1\cr{0}&{0}}\right]}{\left[\matrix{1&0\cr{0}&{0}}\right]})
=0}

Now, using the relations (25), (27),
evaluating all the Chan-Paton's traces
  and denoting for convenience
\eqn\grav{\eqalign{\tau_1\equiv{1}\cr
\tau_2\equiv{e^{{{2i\pi}\over3}}}\cr
\tau_3\equiv{e^{{{4i\pi}\over3}}}\cr
{\vec{p}}={\vec{k}}+{{\vec{b}}\over{2\pi}}}}
 we easily calculate the
first type of string field theory contribution to the
$({\vec{\xi}}{\vec{b}})({\vec{\lambda}}{\vec{\lambda^{*}}})$
coupling  determined by the vertex \eqn\grav{\eqalign{C_3^{(1)}=
{1\over{12g^2}}<<({\hat{Q}}_{brst}{\hat{P}})
{\hat{U}}({\hat{\eta}}_0{\hat{W}})>>-
<<({\hat{Q}}_{brst}{\hat{P}})({\hat{\eta}}_0{\hat{U}})
{\hat{W}}>>\cr= <\xi_q(0)\lambda_s({\vec{k}})\lambda_t^{*}
(-{\vec{k}})<:c\partial{X^q}:(\tau_1)
:c(e^\phi+e^{-3\phi})(1+e^\chi)\psi_0...\psi_3\psi^s
{e^{i{\vec{p}}{\vec{X}}}}:(\tau_2)\cr
:c(e^\phi+e^{-3\phi})(1+e^\chi)\psi_0...\psi_3\psi^s
{e^{-i{\vec{p}}{\vec{X}}}}:(\tau_3)>
\cr={1\over{24\pi}}({\vec{b}},{\vec{\xi}})({\vec{\lambda({\vec{k}})}}
{\vec{\lambda^{*}}})
(\tau_1-\tau_2)(\tau_1-\tau_3)(\tau_2-\tau_3)\times
(\tau_2-\tau_3)^3(\tau_2-\tau_3)^{-5}({2\over{{\tau_1-\tau_2}}}
-{2\over{\tau_1-\tau_3}})\cr
+\lbrace{permut.{\tau_2}\leftrightarrow{\tau_3}}\rbrace
\cr={1\over{6\pi{g^2}}}({\vec{b}},{\vec{\xi}})({\vec{\lambda({\vec{k}})}}
{\vec{\lambda^{*}}})({{{\tau_1}-{\tau_3}}\over{\tau_2-\tau_3}}
-{{{\tau_1}-{\tau_2}}\over{\tau_2-\tau_3}})
={1\over{6\pi{g^2}}}({\vec{b}}{\vec{\xi}})({\vec{\lambda({\vec{k}})}}
{\vec{\lambda^{*}}})}} The second group of three point
correlators contributing to the
$({\vec{b}}{\vec{\xi}})({\vec{\lambda}({\vec{k}})}
{\vec{\lambda^{*}}})$ coupling, that
 involves
the BRST commutators (26)  with the five-form requires much more work
 to calculate; the relevant
non-vanishing correlators contribute to the three-point vertex given by
\eqn\grav{\eqalign{C_3^{(2)}={1\over{12g^2}}
(<<({\hat{Q}}_{brst}{\hat{U}})
{\hat{W}}({\hat{\eta}}_0{\hat{P}})>>-
<<({\hat{Q}}_{brst}{\hat{U}})({\hat{\eta}}_0{\hat{W}})
{\hat{P}}>>\cr+
<<({\hat{Q}}_{brst}{\hat{W}})
{\hat{P}}({\hat{\eta}}_0{\hat{U}})>>-
<<({\hat{Q}}_{brst}{\hat{W}})({\hat{\eta}}_0{\hat{P}})
{\hat{U}}>>)}}
Substituting the string field expansion (23),
using the expressions for BRST commutators derived in
(26) and evaluating the three-point functions
at $\tau_1=1,\tau_2=e^{{{2i\pi}\over3}},\tau_3=e^{{{4i\pi}\over3}}$
as previously,
we find that the second type three-point vertex  gives
\eqn\lowen{C_3^{(2)}= {{8.83272}\over{g^2}}{\times}
({\vec{b}}{\vec{\xi}})({\vec{\lambda}({\vec{k}})}
{\vec{\lambda^{*}}})}
Adding together these two types of vertices, we obtain the answer for
the relevant coupling :
\eqn\lowen{C_3=C_3^{(1)}+C_3^{(2)}= {{8.88577}\over{g^2}}
\times({\vec{b}}{\vec{\xi}})({\vec{\lambda}({\vec{k}})}
{\vec{\lambda^{*}}})}
Now, to normalize $\lambda$ in terms of collective coordinates of the brane
 we have to compare it with the tension change given by
${1\over{2\pi^2}}({\vec{b}}{\vec{Y}})$ of (22) where ${\vec{Y}}$
is a collective coordinate associated with massless photonic
excitation of an open string stretched between the branes. As has
been pointed out in  ~{\szb}, the relation between this
collective coordinate and polarization vector ${\vec{\xi}}$ of
the photon is given by
\eqn\lowen{{\vec{\xi}}=-{{{\vec{Y}}}\over{\pi{\sqrt{2}}}}}
Substituting into (32) we get
\eqn\lowen{C_3=2({\vec{b}}{\vec{Y}})({\vec{\lambda}}{\vec{\lambda^{*}}})}
On the other hand, it follows from (22) that $C_3$ should be given by
\eqn\lowen{C_3={1\over{2\pi^2}}
({\vec{b}}{\vec{Y}})({\vec{\Lambda}}{\vec{\Lambda^{*}}})}
with ${\vec{\Lambda}}$ and ${\vec{\Lambda}^{*}}$ being collective
coordinates of the brane and the anti-brane associated with the
five-forms. Therefore we obtain
\eqn\lowen{{\vec{\lambda}}={1\over{2\pi}}{\vec{\Lambda}}}
Substituting this to the kinetic term (20), (21) of
string field theory Lagrangian, associated with the five-form,
we express it in terms of the brane collective coordinates
and get
\eqn\lowen{
L_{kin}^{(\lambda)}={1\over{4\pi{g^2}}}\sum_k(k^2)({\vec{\Lambda}}
{\vec{\Lambda^{*}}})}
where the coefficient before the sum over $k$ gives us
a half of the brane tension.
Therefore the D-brane tension, associated with the massless five-form,
 is given by
\eqn\lowen{T={1\over{2\pi{g^2}}}}
It is quite remarkable that the massless 5-form reproduces the
well-known tension value of a standard $D$-brane. In the next
sections we will perform the computation of the ghost tachyon
potential and compare this tension with the energy density at the
minimum of the potential to demonstrate the phenomenon of
cancellation.

\medskip

\centerline{\bf 5. BRST Commutators of String Fields}

\medskip

To start the computation of the potential in the OSFT
we will need expressions for commutators of the BRST charge with string
fields in the twist-odd sector, including tachyonic and ghost tachyonic modes.
 The commutators are straightforwardly evaluated and are given by:
\eqn\grav{\eqalign{\lbrace{{\hat{Q}}_{brst}},
{\hat{T}}\rbrace=(-{1\over2}:c\partial{c}:
e^{\chi-\phi}+{1\over4}\gamma)\otimes{I}\cr
{\lbrace{{\hat{Q}}_{brst}},ce^{\chi-3\phi}
\psi_0...\psi_3\rbrace\otimes{\sigma_1}}=(-{1\over2}
:c\partial{c}:e^{\chi-3\phi}\psi_0...\psi_3)\otimes{I}\cr
{\lbrace{Q_{brst}},{\hat{A}}}\rbrace=(-2e^{3\sigma+2\chi-2\phi}
+{1\over2}{\lbrack}c(2\partial^2\phi-2\partial^2\chi\cr+
4(\partial\phi-\partial\chi)^2)+2\partial{c}(2\partial\phi-2\partial\chi)
\rbrack)\otimes{\sigma_1}\cr
{\lbrace{{\hat{Q}}_{brst}},{\hat{E}}\rbrace}=(\partial(c\partial\chi)-{1\over2}
\gamma(\psi_m\partial{X^m})-{1\over2}\gamma^2b)\otimes{\sigma_1}\cr
{\lbrace{Q_{brst}},{\hat{F}}}\rbrace=-{1\over2}:c\partial{c}e^{\chi-\phi}
(\psi_m\partial{X_m})-{1\over8}\gamma(\psi_m\partial{X^m})+
{1\over4}c(\partial{X_m}\partial{X^m}+\psi_m\partial{\psi^m})\cr+
(2\partial^2\phi-2\partial^2\chi+4(\partial\phi-\partial\chi)^2)
\otimes{\sigma_1}\cr
{\lbrace{{\hat{Q}}_{brst}}},ce^{\chi+\phi}\psi_0...\psi_3\otimes{\sigma_1}
\rbrace=(-{1\over2}:c\partial{c}:e^{\chi+\phi}\psi_0...\psi_3\cr
-{1\over2}e^{2\phi}\lbrack\psi_0...\psi_3((\psi_m\partial{X^m})
(\partial\phi-\partial\chi)+\partial(\psi_m\partial{X^m}))\cr
+\psi_{\lbrack{0}}...\psi_2\partial{X_{3\rbrack}}(\partial^2\phi
-\partial^2\chi+(\partial\phi-\partial\chi)^2)\cr+
\psi_{\lbrack{0}}...\psi_2\partial{X_{3\rbrack}}(\partial\phi-\partial\chi)\cr
-{1\over4}e^{3\phi-\chi}\psi_0...\psi_3{P^{(4)}_{2\phi-2\chi-\sigma}})
\otimes{I}\cr
\lbrace{\hat{Q}}_{brst},ce^{\chi-3\phi}\partial{X_a}\partial{X_b}
\otimes{\sigma_1}\rbrace=-{1\over2}:c\partial{c}:e^{\chi-3\phi}
\partial{X_a}\partial{X_b}\otimes{I}\cr
\lbrace{\hat{Q}}_{brst},ce^{\chi+\phi}\partial{X_a}\partial{X_b}
\otimes{\sigma_1}\rbrace=(-{1\over2}:c\partial{c}:e^{\chi+\phi}
\partial{X_a}\partial{X_b}\cr-{1\over2}ce^{2\phi}\lbrace
\partial{X_a}\partial{X_b}((\partial\phi-\partial\chi)(\psi_m\partial{X^m})
+\partial(\psi_m\partial{X^m})\cr
+(\psi_a\partial{X_b}+\psi_b\partial{X_a})P^{(3)}_{\phi-\chi}+
(\partial{X_a}\partial\psi_b+\partial{\psi_a}\partial{X_b})
+\partial^2{X_a}\psi_b+\partial^2{X_b}\psi_a)P^{(2)}_{\phi-\chi}\cr+
{1\over4}e^{3\phi-\chi}\partial{X_a}\partial{X_b}P^{(4)}_{2\phi-2\chi-\sigma}
\otimes{I})}}
where
$P^{(n)}_{f(\phi,\chi,\sigma)}$ is the polynomial of conformal wheight $n$
given by the pre-exponential factor that appears as a result of taking
the $n$th derivative (with respect to z) of the operator
$e^{f(\phi,\chi,\sigma)}$ multiplied by ${1\over{n!}}$
(with f being a linear function
of derivatives of the scalar fields). In particular,
\eqn\grav{\eqalign{P^{(4)}_{2\phi-2\chi-\sigma}=
{1\over{24}}({2}\partial^4\phi-2\partial^4\chi-\partial^4\sigma)
+{1\over6}(2\partial^3\phi-2\partial^3\chi-\partial^3\sigma)
(2\partial\phi-2\partial\chi-\partial\sigma)\cr+
{1\over4}(2\partial^2\phi-2\partial^2\chi-\partial^2\sigma)
(2\partial\phi-2\partial\chi-\partial\sigma)^2+{1\over8}
(2\partial^2\phi-2\partial^2\chi-\partial^2\sigma)^2\cr+{1\over{24}}
(2\partial\phi-2\partial\chi-\partial\sigma)^4}}

In the next section we shall calculate the ghost tachyon
potential using the string field truncated expansion (17), the
modified tension of the $D3$-brane (associated with the 5-form)
and to verify that they cancel each other. We will perform the
computations up to the level
${3\over2}$ for the OSFT modes belonging to $H_{0,0}$ but only up
to the level ${1\over2}$ for
$\hat\Phi\subset{\tilde{H}_{0,-2}}\oplus{\tilde{H}_{0,-2}}$, i.e.
for string field modes associated with the brane-like states.

\medskip

\centerline{\bf 6. Computation of the ghost tachyon potential}

\medskip

In this paragraph we will demonstrate
the computation of the quartic ghost tachyonic term
$\sim{\tau^4}$ and present the answer to the tachyon potential
that we have found (calculations involving the ghost tachyonic
modes and the level $3/2$ $H_{0,0}$ modes are in fact quite
cumbersome and one has to use Mathematica to get numerical
answers  as well as to analyze the extremal points of the
potential). The relevant 4-point functions involve ghost tachyonic
4-forms with both $+1$ and $-3$ bosonic ghost number. However, as
all the vertex operators are multiplied by the $c$ ghost field,
it is clear that only the corelators involving
${\lbrace{{\hat{Q}}_{brst}}},ce^{\chi+\phi}\psi_0...\psi_3$
$\otimes{\sigma_1}\rbrace$ contribute to the potential, since, as
one can see from the table of BRST commutators (39),
 this is the only commutator
containing a term without a c-ghost given by
$:P^{(4)}_{2\phi-2\chi-\sigma}e^{3\phi-\chi}\psi_0...\psi_3$.
Using $:\eta_0{e^{\chi-3\phi}}:\sim{e^{-3\phi}}$,
$:\eta_0{e^{\chi+\phi}}:\sim{e^{\phi}}$  and evaluating the CP
factor trace we find that the relevant contribution
 is given by the following
piece of the OSFT action (5):
\eqn\grav{\eqalign{A_{\tau^4}={{\tau^4}\over{48g^2}}\cr
\lbrace<<:P^{(4)}_{2\phi-2\chi-\sigma}e^{3\phi-\chi}\psi_0...\psi_3
:ce^{-3\phi}\psi_0...\psi_3::ce^{\chi-3\phi}\psi_0...\psi_3:
:ce^{\chi+\phi}\psi_0...\psi_3:>>\cr+
<<:P^{(4)}_{2\phi-2\chi-\sigma}e^{3\phi-\chi}\psi_0...\psi_3
:ce^{-3\phi}\psi_0...\psi_3::ce^{\chi+\phi}\psi_0...\psi_3:
:ce^{\chi-3\phi}\psi_0...\psi_3:>>\cr+
<<:P^{(4)}_{2\phi-2\chi-\sigma}e^{3\phi-\chi}\psi_0...\psi_3
:ce^{\phi}\psi_0...\psi_3::ce^{\chi-3\phi}\psi_0...\psi_3:
:ce^{\chi-3\phi}\psi_0...\psi_3:>>\cr+
<<:P^{(4)}_{2\phi-2\chi-\sigma}e^{3\phi-\chi}\psi_0...\psi_3
y:ce^{\chi-3\phi}\psi_0...\psi_3::ce^{\chi+\phi}\psi_0...\psi_3:
:ce^{-3\phi}\psi_0...\psi_3:>>\cr+
<<:P^{(4)}_{2\phi-2\chi-\sigma}e^{3\phi-\chi}\psi_0...\psi_3
:ce^{\chi-3\phi}\psi_0...\psi_3::ce^{\chi+\phi}\psi_0...\psi_3:
:ce^{-3\phi}\psi_0...\psi_3:>>\cr+
<<:P^{(4)}_{2\phi-2\chi-\sigma}e^{3\phi-\chi}\psi_0...\psi_3
:ce^{\chi+\phi}\psi_0...\psi_3::ce^{\chi-3\phi}\psi_0...\psi_3:
:ce^{-3\phi}\psi_0...\psi_3:>>\cr+
<<:P^{(4)}_{2\phi-2\chi-\sigma}e^{3\phi-\chi}\psi_0...\psi_3
:ce^{\chi-3\phi}\psi_0...\psi_3::ce^{\chi-3\phi}\psi_0...\psi_3:
:ce^{\phi}\psi_0...\psi_3:>>\cr
-2<<:P^{(4)}_{2\phi-2\chi-\sigma}e^{3\phi-\chi}\psi_0...\psi_3
:ce^{\chi+\phi}\psi_0...\psi_3::ce^{-3\phi}\psi_0...\psi_3:
:ce^{\chi-3\phi}\psi_0...\psi_3:>>\cr
-2<<:P^{(4)}_{2\phi-2\chi-\sigma}e^{3\phi-\chi}\psi_0...\psi_3
:ce^{\chi-3\phi}\psi_0...\psi_3::ce^{-3\phi}\psi_0...\psi_3:
:ce^{\chi+\phi}\psi_0...\psi_3:>>\cr
-2<<:P^{(4)}_{2\phi-2\chi-\sigma}e^{3\phi-\chi}\psi_0...\psi_3
:ce^{\chi-3\phi}\psi_0...\psi_3::ce^{\phi}\psi_0...\psi_3:
:ce^{\chi-3\phi}\psi_0...\psi_3:>>\rbrace}} Let us demonstrate,
for example, the calculation of the first correlator,

$<<:P^{(4)}_{2\phi-2\chi-\sigma}e^{3\phi-\chi}\psi_0...\psi_3$
$:ce^{-3\phi}\psi_0...\psi_3::ce^{\chi-3\phi}\psi_0...\psi_3:$
$:ce^{\chi+\phi}\psi_0...\psi_3:>>$, the evaluation of all others
is totally analogous.
Denoting for convenience
\eqn\grav{\eqalign{x\equiv1,\cr
y{\equiv}e^{{{i\pi}\over2}}=i,\cr
z{\equiv{e^{i\pi}}=-1},\cr
a{\equiv}e^{{{3i\pi}\over2}}=-i}}
and using the fact that for primary any dimension $h$ primary operator $V$
\eqn\lowen{f_i^{(N)}{\circ}\lbrace{Q_{brst}},V(0)\rbrace=
\lbrace{Q_{brst}},f_i^{(N)}{\circ}V(0)\rbrace=
{({{df_i^{(N)}(z)}\over{dz}})^h}\lbrace{Q_{brst}},V(f_i^{(N)}(0))\rbrace}
we obtain:
\eqn\grav{\eqalign{<<:P^{(4)}_{2\phi-2\chi-\sigma}e^{3\phi-\chi}\psi_0...\psi_3
:ce^{-3\phi}\psi_0...\psi_3::ce^{\chi-3\phi}\psi_0...\psi_3:
:ce^{\chi+\phi}\psi_0...\psi_3:>>\cr\equiv
<<f_1^{(4)}{\circ}P^{(4)}_{2\phi-2\chi-\sigma}e^{3\phi-\chi}\psi_0...\psi_3(0)
f_{2}^{(4)}{\circ}ce^{-3\phi}\psi_0...\psi_3(0)\cr\times
f_{3}^{(4)}{\circ}ce^{\chi-3\phi}\psi_0...\psi_3(0)
f_{4}^{(4)}{\circ}ce^{\chi+\phi}\psi_0...\psi_3(0)>>\cr=
\prod_{k=1^4}(f_{k}^{(4)}(0))\prime
<P^{(4)}_{2\phi-2\chi-\sigma}e^{3\phi-\chi}\psi_0...\psi_3(x)\cr\times
ce^{-3\phi}\psi_0...\psi_3(y)ce^{\chi-3\phi}\psi_0...\psi_3(z)
ce^{\chi+\phi}\psi_0...\psi_3(a)>}}
Firstly, evaluating the matter part of the correlator
and substituting numerical values of $x,y,z$ and $a$ we get:
\eqn\grav{\eqalign{A_{\tau^4}^{matter}=
<:\psi_0...\psi_3:(x):\psi_0...\psi_3:(y)
:\psi_0...\psi_3:(z):\psi_0...\psi_3:(a)>\cr=
(((x-y)^{-4}(z-a)^{-4}+(x-z)^{-4}(y-a)^{-4}+(x-a)^{-4}(y-z)^{-4})\cr
-4((x-y)^{-3}(z-a)^{-3}((y-z)^{-1}(x-a)^{-1}+(y-a)^{-1}(x-z)^{-1})\cr
+(x-z)^{-3}(y-a)^{-3}((x-y)^{-1}(z-a)^{-1}+(x-a)^{-1}(y-z)^{-1})\cr
+(x-a)^{-3}(y-z)^{-3}((x-z)^{-1}(y-a)^{-1}+(x-y)^{-1}(z-a)^{-1}))\cr
+12((x-y)^{-2}(y-z)^{-2}(z-a)^{-2}(x-a)^{-2}+(x-z)^{-2}(x-a)^{-2}(y-z)^{-2}
(y-a)^{-2}\cr+(x-y)^{-2}(x-a)^{-2}(z-a)^{-2}(y-z)^{-2})\cr
-24((x-y)^{-2}(z-a)^{-2}(y-z)^{-1}(x-z)^{-1}(x-a)^{-1}(y-a)^{-1}\cr+(x-z)^{-2}
(y-a)^{-2}(x-y)^{-1}(x-a)^{-1}(y-z)^{-1}(z-a)^{-1}\cr
+(x-a)^{-2}(y-z)^{-2}(x-y)^{-1}(x-z)^{-1}(y-a)^{-1}(z-a)^{-1}))\cr
=-{{159}\over{64}}}}
To evaluate the ghost correlator involving the exponents and the
$P^{(4)}_{2\phi-2\chi-\sigma}$ polynomial,
 it is convenient to use the O.P.E.'s:
\eqn\grav{\eqalign{:2\partial\phi-2\partial\chi-\partial\sigma:(z):ce^{-3\phi}(w)
\sim{{5}\over{z-w}}e^{-3\phi}(w)\cr
:2\partial\phi-2\partial\chi-\partial\sigma:(z):ce^{\chi-3\phi}(w)
\sim{{3}\over{z-w}}e^{\chi-3\phi}(w)\cr
:2\partial\phi-2\partial\chi-\partial\sigma:(z):ce^{\chi+\phi}(w)
\sim{-}{{5}\over{z-w}}e^{\chi+\phi}(w)}}
Then, using the O.P.E. (26) and the polynomial definition  (20)
it is straightforward to compute the ghost part of the correlator (24)
and the answer is
\eqn\grav{\eqalign{<P^{(4)}_{2\phi-2\chi-\sigma}e^{3\phi-\chi}(x)
ce^{-3\phi}(y)ce^{\chi-3\phi}(z)ce^{\chi+\phi}(a)>\cr
=\lbrack(1/24)(18/((x-z)^4)-30/((x-y)^4)+30/((x-a)^4))+(1/6)(10/((x-y)^3)\cr
+6/((x-z)^3)-10/((x-a)^3))(5/(x-y)+3/(x-z)-5/(x-a))+(1/4)(-3/((x-z)^2)\cr
-5/((x-y)^2)+5/((x-a)^2))(5/(x-y)+3/(x-z)-5/(x-a))^2+(1/8)(-3/((x-z)^2)\cr
-5/((x-y)^2)+5/((x-a)^2))^2+(1/24)(5/(x-y)+3/(x-z)-5/(x-a))^4\rbrack\cr
{\times}(x-y)^9(x-z)^8(x-a)^{-4}(y-z)^-8(y-z)^-8(y-a)^4(z-a)^5}}
Note that all the correlators in (21) differ only in their ghost
parts while their matter part is always the same, given by
$A_{\tau^4}^{matter}$ of (25). Computing
 all the ghost correlators
of (21) similarly to (27), summing them up and multiplying by
the
we get the following answer for the $\tau^4$ contribution:
\eqn\grav{\eqalign{A_{\tau^4}=\prod_{k=1^4}(f_{k}^{(4)}(0))\prime
\times{i\over{g^2}}\lbrace{{256}\over3}({{239}\over8}-{{77}\over2})
+{{145}\over{2048}}\rbrace \times{\tau^4}\cr=
\prod_{k=1^4}(f_{k}^{(4)}(0))\prime\times(-{{735.929i}\over{g^2}})\tau^4}}
Finally, evaluating the conformal transformation factor we
get
\eqn\lowen{{A_{\tau^4}={{735.929i}\over{g^2}}\tau^4}}

This concludes the computation of the $\tau^4$ contribution. The
strategy for computing all other OSFT contributions to the
tachyon potential is totally similar; our final result for the
potential is given by:
\eqn\grav{\eqalign{V(t,a,e,f,\tau,\lambda_{ab})=
-2ae-5f^2+at^2+0.25et^2\cr-4.96aet^2+
0.6654e^2t^2+5.476eft^2-5.82f^2t^2\cr+
0.518et^4-0.27777e^2t^4+0.0826f^2Tr(\lambda^2) -
1583.73et^2Tr(\lambda^2)+47022.1325eTr(\lambda^4)\cr
+0.0113118f^2\tau^2+ 44.598428e{Tr(\lambda^2)}x^2+ \cr
10482.76792e^2{Tr(\lambda^2)}\tau^2+0.350694t^2(2{Tr(\lambda^2)}+\tau^2)+
 36.3293et^2(2Tr(\lambda^2)+x^2)\cr
-97.2121e^2t^2(2Tr(\lambda^2)+\tau^2)+
0.25(t^2 +2Tr(\lambda^2)+\tau^2)\cr-735.92919921875\times
(0.515625{Tr(\lambda^2)}^2+1.5Tr(\lambda^4)
+ 0.1015625Tr(\lambda^2)\tau^2 -\tau^4)\cr
+0.8(-63639.61875e{Tr(\lambda^2)}^2
-127279.2375eTr(\lambda^4)-79718.8925e\tau^4)\cr-
 0.1*(-6.7155426\times{10}^6e^2u^2-4.623888\times{10^6}e^2Tr(\lambda^4)\cr-
-6a(\tau^2+2Tr(\lambda^2))+1.4522886\times{10^6}e^2\tau^4)}}
The minimum of this potential is attained at:
\eqn\grav{\eqalign{a=9.5307,f=0.0315,t=0.9148,e=0.136,\cr
\tau=0.0115, Tr(\lambda^2)=0.017, Tr(\lambda^4)=0.00059}}
and the value of the potential at this point is
\eqn\lowen{V_{min}=-{{1.157\times{10^{-1}}}\over{g^2}}}
On the other hand, the D3-brane tension
associated with massless 5-form brane-like state,
 calculated in the appendix,
is given by ${1\over{2\pi{g^2}}}$
so we see that already at the level $3/2$
of the OSFT about $99.7$ percent of the brane tension is cancelled
by the ghost tachyon potential! In particular
this may explain why the usual perturbative string theory
does not feel the brane-like states and why the latter do not cause
any difficulties with unitarity: it appears that these states are simply
``screened'' by the tachyonic veil,
associated with brane-like tachyonic states (2), (3).

\medskip

\centerline{\bf 7. Extra Time Dimensions, Brane phenomenology and
Tachyon Condensation}

\medskip

In this section we shall discuss possible implications of the
above results for phenomenological brane models with extra
dimensions ~{\rsh}. The great disadvantage of known brane world
scenarios ~{\bran} is that they essentially include objects with
negative tensions, antibranes which in reality are unstable and
do not exist. In a separate development, it has been shown
recently that when one places the brane-antibrane pair in a de
Sitter-type background with time-like extra dimension, tachyonic
modes appear among solutions to effective gravity equations of
motion in such a background ~{\ck} In this section we shall
attempt to show that these two instabilities, i.e. the negative
tension antibrane  cancel each other by the mechanism very
similar to the ghost tachyon condensation discussed above.
Namely, we shall argue that, in terms of string theory,  the
tachyonic modes corresponding to the e.o.m. solutions correspond
precisely to the ghost tachyonic vertex operators described
above. As a result, one can again use the OSFT formalism to show
that these tachyonic modes again have an effective potential
bounded from below and the energy density at the potential
minimum precisely compensates the antibrane tension. The analysis
is quite similar to the one performed in the case of
 single $D$-brane; the only difference is that now the usual
matter tachyonic field needs to be switched off, i.e. the ghost
tachyon potential (30) must be analyzed at $t=0$.

The tachyonic potential (30) with the extra condition
$t=0$ reaches its minimum at
\eqn\grav{\eqalign{f=0.0549,e=0.258,\cr
\tau=1.0035\times{10^{-7}}, Tr(\lambda^2)=0.0141, Tr(\lambda^4)=0.00087}}
and the potential value at the extremum is given by
\eqn\lowen{V_{min}={{1.139\times{10^{-1}}}\over{g^2}}} On the
other hand the antibrane tension is given by
-${1\over{2\pi{g^2}}}$, i.e. in this case the ghost tachyon
potential compensates about $97$ percent of the antibrane tension.

\medskip

\centerline{\bf 8. Discussion}

\medskip

It appears that there is a deep physical reason behind relation
of the tachyonic modes in the de Sitter-type backgrounds and
ghost tachyonic vertex operators. Some  time ago  in an unrelated
development ~{\myself} it has been shown that the backgrounds of
such a type can be understood as a result of dynamical
compactification caused by the massless 5-form brane-like state.
To explore the mechanism of the dynamical compactification of
flat ten-dimensional space-time  on $AdS_5\times{S^5}$ due to
presence of the $V_5$  vertex in the
 sigma-model one has to study the  modification of the dilaton's
beta-function  in the $V_5$-background. Such an analysis has been
carried out in ~{\myself}.
 The analysis of the dilaton's beta-function
shows that the compactification
on $AdS_5\times{S^5}$ occurs
as a result of  certain very special non-Markovian stochastic process.
 Namely,
 the $V_5$ background in the sigma-model
 has a meaning of a ``random force'' term with the $V_5$-operator
playing the role of a non-Markovian stochastic noise,
which correlations are determined
by the worldsheet  beta-function associated with the $V_5$ vertex.
The straightforward computation shows that
the dilaton's  beta-function equation
in the presence of the $V_5$-term
has the form of the non-Markovian Langevin equation:

\eqn\lowen{
{{d\varphi(p)}\over{d(log\Lambda)}}=
-\int{d^{10}q}C_\varphi(q)\varphi({{p-q}\over2})\varphi({{p+q}\over2})
+\eta_{5}(p^{||},\Lambda)}
where
\eqn\lowen{\eta_{5}(p^{||},\Lambda)\equiv
-\lambda_0^2
(1+\lambda_0\int{d^4k_2^{||}}
\int_0^{2\pi}d\alpha\int_0^\infty{dr}rV_5(r+\Lambda,\alpha,k^{||}))}
In this equation
 the role of the stochastic noise term being played by
the truncated worldsheet integral of the $V_5$-vertex. The
logarithm of the worldsheet  cutoff parameter plays the role of
the stochastic time in the Langevin equation for stochastic
quantization ~{\parisi, \huffel, \periwal}. The noise is
non-Markovian and it is generated by
 the $V_5$ operator , as was already noted above.

The noise correlations in stochastic time are given by the
worldsheet correlators of the $V_5$ vertices (one has to take
their worldsheet integrals at different cutoff values and to
compute  and to evaluate the cutoff dependence). Knowing the
$V_5$-noise correlators it is then straightforward to derive the
corresponding non-Markovian Fokker-Planck equation for this
stochastic process and to show that the Fokker-Planck
distribution solving this equation is given by
 the exponent of the ADM-type  $AdS_5$ gravity Hamiltonian
(computed from the $AdS_5$ gravity action at a constant radial
$AdS$ ``time'' slice using the Verlinde's prescription
~{\verlinde, \verlind}. Such a mechanism naturally relates the
radial $AdS$ coordinate, stochastic time and the worldsheet
cutoff, pointing out an intriguing relation between holography
principle, AdS/CFT correspondence ~{\ampf, \ampr, \malda, \wit}
and non-Markovian stochastic processes.
 Therefore from space-time point of view the
$V_5$ insertion leads to non-Markovian stochastic process
which deforms flat  space-time
geometry  the one of  $AdS$ with the role of the radial $AdS$
coordinate played by the stochastic time parameter.
At the same time, if one studies the
 gravity equations of motion  in a background
in which the radial $AdS$ dimension is replaced by the time-like
coordinate one finds a set of tachyonic solutions to these
equations, including tachyonic scalars and tachyonic gravitons.
Given these tachyons in the and observing that the AdS background
can be represented by insertion related to the $5$-form vertex
operators (1), the appearance of tachyons in higher ghost number
cohomologies ~{\myselff} is not incidental. It appears that they
merely reflect the  $AdS$ structure (with the time-like radial
direction) of the new space-time geometry created by the
massless  $5$-form. On the other hand, due to the demonstrated
analogue of Sen's mechanism of ghost tachyon condensation, the
massless $5$-form (1) and the tachyons (2),(3) ``compensate''
each other and this is why the conventional perturbative NSR
string theory does not ``see'' neither these states nor the $AdS$
space-time geometry.

\medskip

\centerline{\bf Acknowledgements} 

\medskip

D.P. wishes to gratefully
acknowledge the support and the hospitality of High Energy
Accelerator Research Organization (KEK), Tsukuba, Japan and
particularly N. Ishibashi and Y. Kitazawa. This work was
partially supported by the Academy of Finland Project No. 163394.

\listrefs
\end